\documentclass[useAMS,usenatbib]{mn2e}

\DeclareSymbolFont{cmletters}{OML}{cmm}{m}{it}
\DeclareMathSymbol{v}{\mathalpha}{cmletters}{"76}

\usepackage{tabularx,ragged2e,booktabs,caption}
\usepackage{amsmath}
\usepackage{amssymb}
\usepackage{epsfig}
\usepackage{graphicx}
\usepackage{ifthen}
\usepackage{latexsym}
\usepackage{rotating}
\usepackage{subfigure}
\usepackage{times,epsf}
\usepackage{txfonts}
\usepackage{varioref}
\usepackage{verbatim}
\usepackage{array}
\usepackage{url}
\usepackage{color}
\usepackage[dvipsnames]{xcolor}
\usepackage[T1]{fontenc}

\newcommand{\be}{\begin{equation}}
\newcommand{\ee}{\end{equation}}
\newcommand{\bea}{\begin{eqnarray}}
\newcommand{\eea}{\end{eqnarray}}

\newcommand\apj{Astrophysical Journal}
\newcommand\apjl{Astrophysical Journal Letters}

\newcommand\aap{Astronomy \& Astrophysics}

\newcommand\mnras{Monthly Notices of the Royal Astronomical Society}

\newcommand\pasj{Publications of the Astronomical Society of Japan}

\newcommand\jqsrt{Journal of Quantitative Spectroscopy and Radiative Transfer}

\newcommand{\der}[2]{\frac{{\rm d}#1}{{\rm d}#2}}

\newcommand{\Medd}{\dot M_{\rm Edd}}
\newcommand{\rg}{r_{\rm g}}
\newcommand{\tg}{t_{\rm g}}

\title[Magnetically supported thin accretion disks]{Thin accretion disks are stabilized by
  a strong magnetic field}
\author[A. S\k{a}dowski]
       {Aleksander S\k{a}dowski$^{1,2}$\thanks{E-mail: asadowsk@mit.edu (AS)} \\
        $^1$ MIT Kavli Institute for Astrophysics and Space Research
77 Massachusetts Ave, Cambridge, MA 02139, USA\\
$^2$ Einstein Fellow}

\begin{document}

\maketitle

\label{firstpage}

\begin{abstract}
  By studying three-dimensional, radiative, global simulations of
  sub-Eddington, geometrically thin black hole accretion flows we show
  that thin disks which are dominated by magnetic pressure are stable
  against thermal instability. Such disks are thicker than predicted
  by the standard model and show significant amount of dissipation
  inside the marginally stable orbit. Radiation released in this
  region, however, does not escape to infinity but is advected into
  the black hole. We find that the resulting accretion efficiency
  ($5.5\pm0.5\%$ for the simulated $0.8\dot M_{\rm Edd}$ disk) is very
  close to the predicted by the standard model ($5.7\%$).
\end{abstract}


\section{Introduction}
\label{s.introduction}

Most of the Galactic black hole (BH) X-ray binaries cycle through outbursts and
quiescent states as a result of modulation of the accretion rate by
ionization instability in the outer disk regions \citep{lasota+instability,coriat+12}. During the
outbursts they show luminosities of the order of $1-30\%$ of the
Eddington luminosity, $L_{\rm Edd}$ \citep{maccarone+03,mcclintock+06}. According to the standard disk
theory \citep{ss73}, such luminosities correspond to the radiation-pressure
dominated, radiatively efficient disk state. X-ray binaries can
remain in such a configuration for months, i.e., for the time 
much longer than the relevant dynamical, thermal, or even viscous
timescales. However, radiation-pressure dominated thin disks are
known to be viscously \citep{lightmaneardley-74} and thermally
\citep{ss76} unstable. 
% The latter instability results from the heating rate
% (proportional to the integrated pressure) responding stronger
% to the perturbations of the equilibrium state than the cooling rate
% (proportional to the midplane pressure). 
If such instabilities
operate, one should expect global limit
cycle bahavior \citep{lasotapelat-91,szuszkiewiczmiller-98}\footnote{Which may be resposible
  for some of the variability patterns of the outliers in the X-ray binary set
  -- GRS 1915+105 and IGR J17091-3624.}, instead of the observed
quasi-steady high/soft
state of most BH X-ray binaries. The disagreement strongly
suggests that our understanding of the physics of thin disk accretion
is not satisfactory.

Accretion flows are known to be turbulent. Because of this fact the
value of analytical
modeling is limited and numerical simulations are required for better
understanding them. Recently,
significant progress has been made through
development of sophisticated magnetohydrodynamical (MHD) and radiation
MHD (RMHD) codes capable of simulating accretion flows both in the shearing
box approximation and in the global context. The application of the latter has so far
been limited to studying geometrically thick accretion disks
\citep[e.g.,][]{ohsuga11,sadowski+dynamo,mckinney+harmrad,jiang+3dsim} which are
known (the optically thick ones) to be stabilized by advection-related
cooling \citep{abra88}. The only insight into the physics of geometrically thin,
radiation-pressure dominated and presumably unstable accretion flows
has come so far from shearing box simulations. Most recently,
\cite{jiang+stability} studied such systems (with zero net magnetic
flux) using a sophisticated
radiation MHD algorithm and have shown that they are indeed unstable,
in disagreement with most of the observed systems.

What does stabilize the astrophysical thin disks? A wide range of ideas has been
proposed, including stabilization by stochastic variability
\citep{janiukmisra-12}, intrinsic delay between heating and cooling
\citep{hirose+09,ciesielski+12}, and magnetic fields. The latter seem
to be most promising as magnetic fields are intrinsically involved in every
accretion event (they make the disks turbulent), and probably are also crucial
in explaining the observed nature of state
transitions \citep{begelmanarmitage-14}.

\cite{begelmanpringle-07} claimed that optically thick, geometrically
thin accretion disks with strong toroidal magnetic field are stable
against thermal and viscous instabilities. The authors based on the
assumption that the field strength for a thin disk saturates at the
level derived by \cite{pessahpsaltis-05}, i.e., when the Alfven speed
roughly equals the geometric mean of the Keplerian speed and the speed
of sound in gas. We do not find this condition satisfied in the
simulated thin disks presented in this work. \cite{oda+09} discussed the stabilizing effect
of strong toroidal magnetic field on thermal stability of optically
thin and thick accretion disks using analytical approach and an ad
hoc, although physically motivated, prescription for the radial
distribution of the magnetic flux. Our work essentially validates
their assumptions and conclusions. Recently, \cite{libegelman-14} have shown that magnetic
fields may help stabilize the disk also through magnetically
driven outflows which decrease the disk temperature and thus help the
disk become more stable at a given accretion rate.

In this work we show that indeed, magnetically supported thin disks
are stable against thermal instability. We perform two
three-dimensional global simulations of radiative, sub-Eddington accretion
flows using a state-of-the-art general relativistic RMHD code. The
simulated accretion flows led to strongly and weakly magnetized states. Only the
former reached thermal equilibrium. The latter showed significant
and consistent excess of cooling.
Our work should be considered as a proof of principle. More
numerical effort is necessary to understand the properties of the
magnetically supported accretion disks accross the spectrum of
accretion rates and large scale properties of  magnetic field.

Our work is organized as follows. In Section~\ref{s.magnstable} we
show with a simple argument that magnetically supported disks are
thermally stable. In Section~\ref{s.method} we present the numerical
methods we applied and the initial conditions for the simulations we
performed. In Section~\ref{s.accretion} we discuss the differences in
the ultimate magnetic field configuration. In
Section~\ref{s.heatingcooling} we compute and compare rates of heating
and cooling. The collapse of the weakly magnetized disk is discussed
in Section~\ref{s.collapse}. The properties of the stable solution are
presented in Sections~\ref{s.properties} and
\ref{s.energetics}, and are followed by Discussion and Summary.

In this work we adopt the following definition
for the Eddington mass accretion rate,
\be
\label{e.medd}
\Medd = \frac{L_{\rm Edd}}{\eta c^2},
\ee
where $L_{\rm Edd}=1.25 \times 10^{38}  M/M_{\odot}\,\rm ergs/s$ is the 
Eddington luminosity, and $\eta$ is the radiative efficiency of a thin
disk around a black hole with a given spin $a_* \equiv a/M$. For zero
BH spin, $\eta\approx 0.057$ and
$\Medd = 2.48 \times 10^{18}M/M_{\odot}  \,\rm g/s$.
Hereafter, we also use the
gravitational radius $r_{\rm g}=GM/c^2$ as the unit of length, and
$t_{\rm g}=r_{\rm g}/c$
as the unit of time.

\section{Magnetic pressure makes thin disks thermally stable}
\label{s.magnstable}

Radiation pressure-dominated disks are known to exhibit thermal
instability coming from the fact that viscous heating responds more
rapidly to changes in midplane pressure than the radiative
cooling \citep{pringle-76,piran-78}. The former is given through,
\be
Q^+=\alpha P_{\rm tot} \der{\Omega_{\rm K}}{r}\approx \alpha p_{\rm
  rad} h\der{\Omega}{r}
\ee
where $\alpha$ is the standard viscosity parameter, $\Omega_{\rm K}$
is the Keplerian angular velocity, $h$ is disk thickness, and $P$ and $p$ denote vertically
integrated and central pressures, respectively. Taking into account the
equation of vertical equilibrium,
\be
\Omega_{\rm K}^2h^2=\frac {P_{\rm tot}}\Sigma= \frac {p_{\rm rad}h}\Sigma,
\ee
one gets,
\be
Q^+= \alpha p_{\rm
  rad}^2 \frac {\Sigma}{\Omega_{\rm K}^2} \der{\Omega}{r}.
\ee
Therefore, in the case of a radiation pressure dominated disk, the
viscous heating depends on the central pressure, here - radiation
pressure, as,
\be
\left.\der{\log Q^+}{\log p_{\rm
  rad}}\right|_{\Sigma} = 2.
\label{e.dqplus}
\ee

Radiative cooling in the diffusive approximation is given through,
\be
Q^-= \frac{p_{\rm
  rad}}{\Sigma \kappa},
\label{e.qminus}
\ee
with $\kappa$ being the opacity coefficient (assumed constant), and depends on the
central pressure as,
\be
\left.\der{\log Q^-}{\log p_{\rm
  rad}}\right|_{\Sigma} = 1.
\label{e.dqminus}
\ee
Weaker slope means that whenever the central pressure gets perturbed, the
heating responds stronger what causes either runaway heating or
cooling \citep[for numerical example of such behavior
see][]{jiang+stability}.

Let us now introduce the
magnetization parameter $\beta'=p_{\rm mag}/p_{\rm tot}$, equal to the
ratio of magnetic to total pressures, and let us neglect the gas
thermal pressure. The rate of radiative cooling is
insensitive to the magnetic pressure and Eqs.~\ref{e.qminus} and
\ref{e.dqminus} hold. The heating rate, however, now equals,
\be
Q^+=\alpha P_{\rm tot} \der{\Omega_{\rm K}}{r}= \frac {\alpha p_{\rm
  rad}^2}{(1-\beta')^2} \frac {\Sigma}{\Omega_{\rm K}^2}\der{\Omega}{r}.
\ee
Taking into account that,
\be
\left.\der{\beta'}{p_{\rm
  rad}}\right|_{p_{\rm mag}}=-\frac{\beta(1-\beta)}{p_{\rm rad}},
\ee
we get,
\be
\left.\der{\log Q^+}{\log p_{\rm
  rad}}\right|_{\Sigma,p_{\rm mag}} = 2(1-\beta'),
\label{e.dQdp}
\ee
what, in the limit of insignificant magnetic field ($\beta'=0$)
recovers the standard result (Eq.~\ref{e.dqplus}). However, if only
the magnetic field dominates the pressure budget, i.e., when
$\beta'>1/2$, the heating rate does no longer increase more rapidly
with the central radiation pressure than the cooling rate and the disk
stabilizes itself. In other words, change of the radiation pressure at
the midplane (resulting, e.g., from gas changing its temperature) does
no longer efficiently translate into the change of the total integrated pressure
(which determines the rate of heating).

\section{Numerical methods}
\label{s.method}

We use general relativistic (GR) radiation magnetohydrodynamical
(RMHD) code \texttt{KORAL} \citep{sadowski+koral,sadowski+koral2}. It
evolves gas, magnetic field and radiation field in a fixed
spacetime described by an arbitrary metric. Magnetic fields are
evolved under the assumption of ideal MHD, i.e., we assume that the
electric field vanishes in the gas comoving frame, and there is no explicit
resistive term. The gas is coupled to radiation through the radiation
four-force describing the energy and momentum exchange through
absorption and scattering. We use the basic thermal
Comptonization as described in \cite{sadowski+dynamo}. We evolve four
quantities describing the radiation field, corresponding to the
radiative energy density
and momentum. The radiation
stress-energy tensor is closed using the M1 closure
\citep{levermore84, sadowski+koral}. In addition, we use the radiative
viscosity prescription given in the Appendix of
\cite{sadowski+dynamo}. We adopted the following formulae for the
absorption and scattering opacities, $\kappa_{\rm abs}$ and
$\kappa_{\rm es}$, respectively,
\bea
\hspace{1cm}\kappa_{\rm abs}&=&6.4\times 10^{22} \rho T^{-7/2}\,\rm cm^2/g,\\
\hspace{1cm}\kappa_{\rm es}&=&0.34 \,\rm cm^2/g.
\eea

\noindent We adopt the Kerr-Shield metric and the
radial cells are distributed exponentially in radius, with the
innermost five cells located inside the BH horizon. The polar cells are
concentrated towards the equatorial plane to provide higher resolution
in the disk region (the effective vertical resolution for simulations
presented here, calculated basing on the vertical cell size at the
equatorial plane, is $N_{\theta,\rm eff}=1250$). The azimuthal cells are distributed uniformly. 

\subsection{Initial state and simulation sequence}
\label{s.initial}

Simulating radiation pressure dominated, geometrically thin accretion
disks requires careful choice of initial conditions. One can easily
imagine that if the initial state is too
far from the equilibrium solution, one may ``overshoot'' and instead
of finding the intended solution at the middle, presumably unstable branch of the
temperature-surface density diagram \citep[the so-called
S-curve,][]{abra88}, one may find a solution at the lower, gas-pressure dominated
branch corresponding to lower accretion rates, or at the upper,
advection-dominated branch. 

The standard way of initializing simulations of accretion disks is to
start from an equilibrium torus of gas threaded by weak poloidal
magnetic field. As a result of magnetorotational instability
\citep[MRI, ][]{balbushawley-mri} this magnetic field triggers
turbulence  and the related dissipative heating. Using the
same setup for simulating thin disks may be problematic. Firstly,
the initial torus may contaminate the most interesting inner region of
the disk by, e.g., blocking radiation coming from that region. Secondly, if
one starts from a thin, radiation pressure supported torus of low
optical depth, the torus may cool significantly (on the
thermal timescale, $t_{\rm th}\approx 1/\alpha \Omega$) by the
time the turbulence triggers in (on the dynamical timescale, $t_{\rm
  dyn}\approx 1/\Omega$), especially for large viscosity $\alpha$
parameters, typical for inner regions of accretion \citep{penna+alpha}.

We decided to follow a different approach. We start from a relatively
optically thick torus
located far from the BH. We evolve
that torus in axisymmetry to let the gas reach the innermost region
and start crossing the BH horizon. Then, the gas, magnetic field and
radiation are rewritten onto a three-dimensional grid. The simulation
is continued in three-dimensions and gives mildly super-Eddington
solution. Then, in the second step, the gas density and magnetic field
are scaled down by constant factors keeping fixed the gas temperature
and magnetic to gas pressure ratio. As a result, we start the ultimate
simulation from a state which is already turbulent and presumably
close to the equilibrium corresponding to a sub-Eddington accretion flow.

We have simulated three models, the parameters of which are specified in
Table~\ref{t.models}. Two fiducial simulations (\texttt{Q} and
\texttt{D}) were initiated with the same torus but with different configurations of magnetic
field (discussed below) and both were evolved in three dimensions 
with the azimuthal wedge limited, for the sake of computational cost,
to $\pi/2$. The third simulation, \texttt{Q}$(2\pi)$, was performed with the full $2\pi$
wedge, ran for a limited time, and have shown very similar properties
to simulation \texttt{Q}. For this reason, we will compare only
simulations \texttt{Q} and
\texttt{D} below.

\begin{table}
\begin{center}
\caption{Model parameters}
\label{t.models}
\begin{tabular}{lccccc}
\hline
\hline
Name 
& $\dot M_{\rm BH}/\dot M_{\rm Edd}$ &
$\Delta t$ &  $\Delta\phi$ & 
$N_r\times N_\theta \times N_\phi$ & $B_{\rm init}$\\
\hline
\texttt{Q} & $0.82$ & $20000\tg$ &
$\pi/2$ & $320 \times 320 \times 32$ & quad.\\
\texttt{Q($2\pi$)}& $0.80$ & $5000\tg$ &
$2\pi$ & $320 \times 320 \times 128$ & quad.\\
\texttt{D} & $1\searrow$ & $9200\tg$ &
$\pi/2$ & $320 \times 320 \times 32$ & dipolar\\
\hline
\hline
\multicolumn{6}{l}{$\dot M_{\rm BH}$ - average accretion rate through the BH horizon,}\\
\multicolumn{6}{l}{$\Delta t$ - duration of the simiulation, $\Delta
  \phi$ - wedge in azimuth,}\\
\multicolumn{6}{l}{$N_r\times N_\theta \times N_\phi$ - resolution in
  radius, polar and azimuthal angles,}\\
\multicolumn{6}{l}{$B_{\rm init}$ - configuration of the initial
  magnetic field.}\\
\multicolumn{6}{l}{Other parameters: $M_{\rm BH}=10M_\odot$,
  $a_*=0$, $r_{\rm in}=1.85$,$r_{\rm out}=500$,}\\
\multicolumn{6}{l}{effective resolution in
  polar angle at the equatorial plane, $N_{\theta,\,\rm eff}=1250$. }\\
\end{tabular}
\end{center}
\end{table}

\subsection{Initial magnetic field}

In Section~\ref{s.magnstable} we argued that a thin, radiatively
efficient accretion disk with radiation pressure dominating over gas
pressure is thermally stable if magnetic pressure overcomes the
radiation pressure and provides most of the pressure
support. Such conditions are not satisfied in the standard
saturation state of magnetorotational instability (MRI). Sophisticated
simulations of shearing boxes have shown that the typical
magnetic to total pressure ratio is only $10\%$ \citep[e.g.,][]{davis+10,shi+10}. This
number increases for simulations with non-zero net vertical magnetic
flux. For sufficiently strong fluxes the entire shearing box may
become magnetically dominated \citep{xuening+13}.

Simulations of accretion flows in the shearing box approximation do not allow for
radial advection of heat or magnetic field. They also neglect the
curvature terms (the characteristic radius is infinity). What is more,
one cannot start a simulation with a non-zero radial flux of magnetic
field -- the shear converts the radial component of magnetic field into azimuthal one and
leads to unlimited grow of the magnetic field. Such a behavior,
however, does not have to occur in a global simulation, where the growing
magnetic field can be advected or escape from the disk.

\begin{figure*}
 \includegraphics[width=1.025\columnwidth]{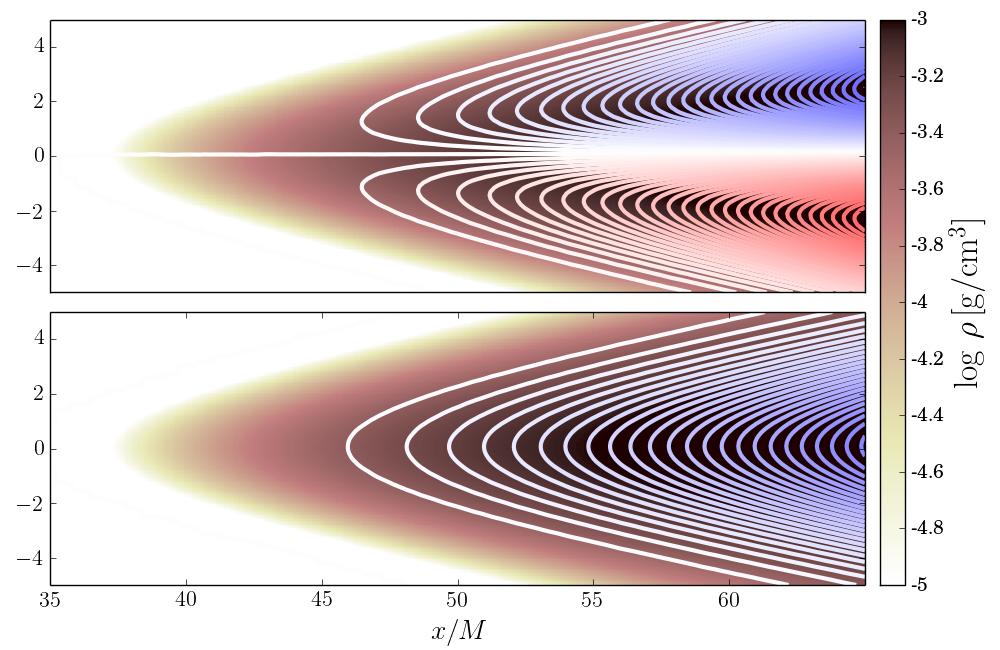}\vspace{-.2cm}
 \includegraphics[width=1.025\columnwidth]{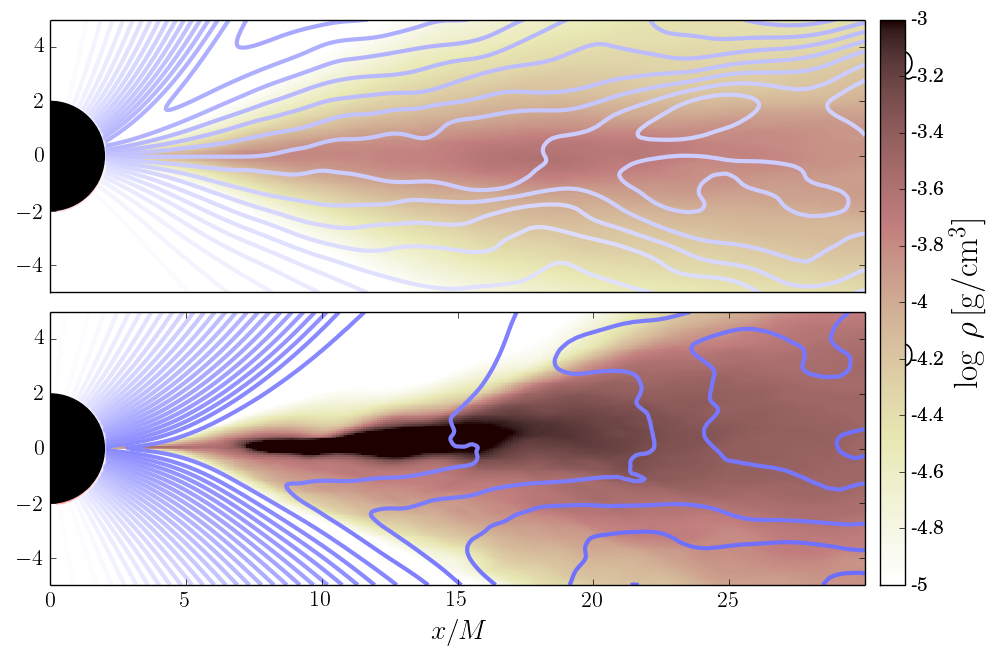}
 \caption{Left panels: Initial configuration of density and the magnetic field in
   simulations \texttt{Q} (top) and \texttt{D} (bottom panel). Colors
   denote clock- (red) and anti-clockwise (blue) loops. Right panels: Magnetic
   field averaged over time and azimuth for the same simulations
   corresponding respectively to $t=5000\div20000 \tg$ and
   $t=2500\div7500 \tg$. Simulation \texttt{Q} (top sub-panel) shows
   much more significant radial net flux of magnetic field which
   results in stronger magnetization.
  }
 \label{f.Bfield}
\end{figure*}

In this work we aim at comparing properties of strongly and weakly
magnetized thin accretion disks.
Having in mind that the large scale configuration of the magnetic
field has an impact on the level of magnetic field saturation, we
decided to test two initial topologies of the magnetic field shown in
the left panels of Fig.~\ref{f.Bfield}. The top sub-panel shows the
initial quadrupolar configuration for models \texttt{Q} and \texttt{Q}$(2\pi)$. The bottom
sub-panel shows the dipolar topology used to initialize model
\texttt{D}. In both cases the magnetic field was normalized not to
exceed $p_{\rm mag}/p_{\rm gas}=1/20$ in the domain. We stress here
that little is known about large scale magnetic fields in accretion
flows and both configurations must be somewhat arbitrary.

 In the global
sense both topologies have zero net-flux of magnetic field - the
loops are contained in the torus and in the computational
domain. However, as accretion proceeds, one may expect that portions
of the torus located at different radii will approximately proceed in
order towards the
BH. According to this picture, the quadrupolar topology would
correspond to zero net vertical flux, in contrast to the dipolar
topology where the vertical field does not cancel out at given
radius.

 For both configurations the net radial flux at given radius is
zero. However, the two are not identical in this aspect. The
quadrupolar topology is asymmetric -- the positive radial magnetic
field is located near the equatorial plane and the negative field
near the torus surface. If only the latter escapes the bulk
of the disk, e.g., buoyantly or dragged by outflowing gas, then one may expect
non-zero radial net flux left in the disk. Similar effect would take
place if the clock- and counter-clockwise loops are not advected in
parallel -- one of the loops may dominate the inner region but the
presence of the other one would make the first asymmetric with respect
to the equatorial plane.

\section{Accretion and ultimate magnetic field}
\label{s.accretion}

\begin{figure}
 \includegraphics[width=1.0\columnwidth]{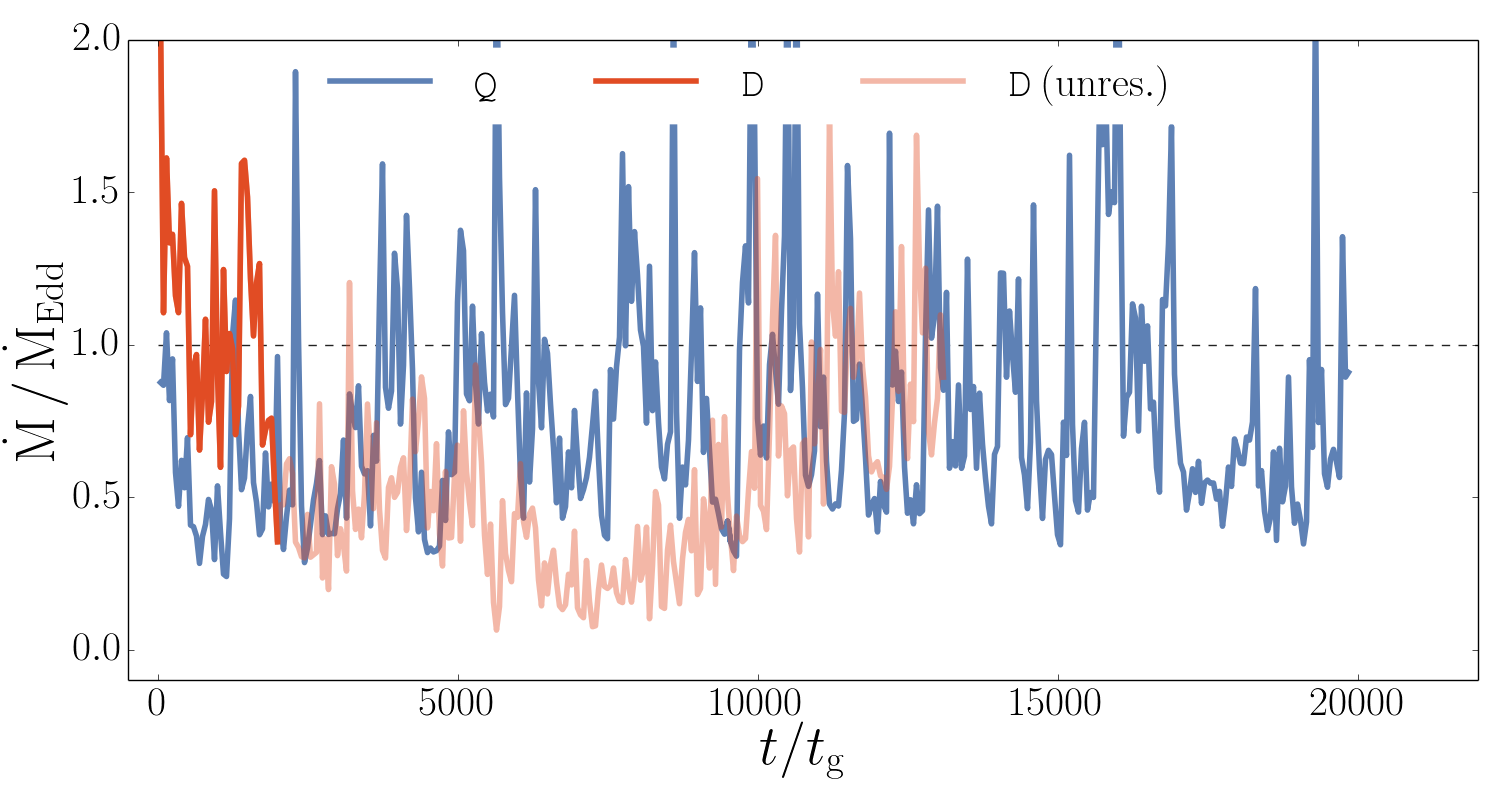}
  \caption{The history of accretion rate through the BH horizon for
    the thin disk simulations. The weakly magnetized simulation
    \texttt{D} became under-resolved after ca. $2000\tg$.
  }
 \label{f.mdots}
\end{figure}

Figure~\ref{f.mdots} shows the history of the accretion rate
through the BH horizon for the simulated models. Simulation \texttt{Q}
(blue line) accreted at relatively steady state with average
accretion rate of ca. $0.8\Medd$. The inflow/outflow
equilibrium, i.e., the region where the accretion rate is constant,
extended in this case to $r\approx15\rg$. The other simulation,
\texttt{D} (red line), did
not show a steady state -- the accretion rate fell down initially from
$\sim 1\Medd$ to $\sim 0.2\Medd$  as a result of unbalanced cooling of
the disk. The related collapse, discussed below, led to
under-resolving the MRI near the equatorial plane. The initial
decrease of the accretion rate was followed by an increase and
return to the original rate $\sim \Medd$. However, the
under-resolved equatorial region makes this stage of the simulation unreliable.

Right panels of Fig.~\ref{f.Bfield} show the average configuration of
the magnetic field in the saturated state of simulation \texttt{Q}
(top) and in the decline phase of simulation
\texttt{D} (bottom sub-panel). There is a qualitative difference between the two. The latter,
initiated with a dipolar magnetic field, led to magnetic field
roughly symmetric with respect to the equatorial plane. Some amount of
magnetic flux has accumulated at the BH and the bulk of the disk is
threaded by mostly vertical net magnetic field. There is no
significant radial component present. 

It is not the case for the simulation initiated with the quadrupolar
field (\texttt{Q}, top sub-panel). One of the set of loops (the
counter-clockwise) preceeded the other and dominated the innermost
region. However, it did not form a symmetric configuration like in
the other case -- the positive radial magnetic field covers larger
volume and, in particular, dominates in the bulk of the disk. The
magnetic pressure corresponding to the net radial component amounts to
$\sim 0.0015$ of the total pressure at radius $\sim 15\rg$.

\begin{figure}
 \includegraphics[width=1.0\columnwidth]{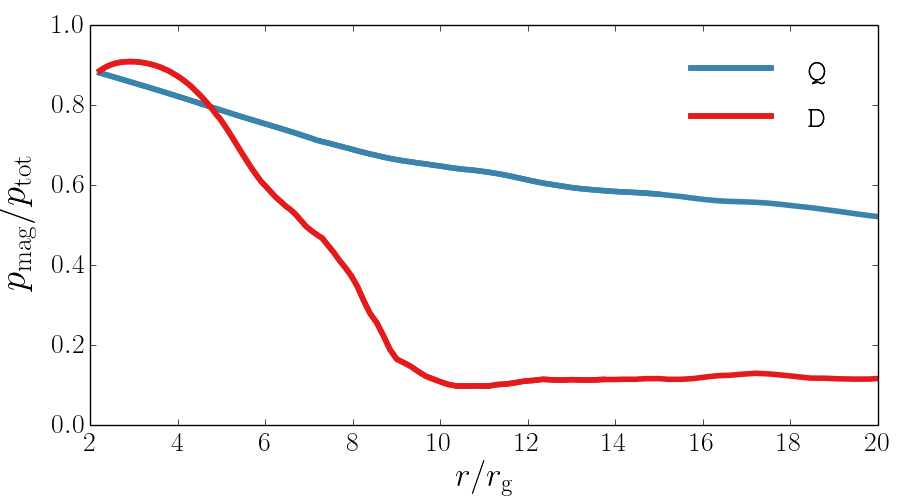}
 \caption{
The ratio of magnetic to total pressures averaged within the disk density
scaleheight for models \texttt{Q} and \texttt{D} corresponding respectively to $t=5000\div20000 \tg$ and
   $t=2500\div7500 \tg$. Significant net radial magnetic flux in the disk results in magnetically supported accretion. 
  }
 \label{f.betas}
\end{figure}

As discussed before, non-zero radial magnetic flux in a shearing flow
leads to a rapid grow of the azimuthal component and, as a result, of the
magnetic pressure. Fig.~\ref{f.betas} compares the magnetic pressure
contribution to the total pressure between the simulations. Simulation
\texttt{D}, which did not develop significant radial flux, shows
magnetic pressure contributing to ca. $10\%$ of the total pressure for
$r\gtrsim10\rg$, in
agreement with the standard MRI saturation state. The other model,
however, the one with non-zero net radial flux of magnetic field in the
disk, has much different properties -- the magnetic pressure grows so
much that it dominates the pressure budget. Near the edge of the
inflow/outflow equilibrium ($r\approx 15\rg$) the magnetic field provides
$60\%$ of the total pressure. The quadrupolar initial
configuration of magnetic field led to the magnetically supported disk
state \citep[compare][]{machida+06}.

\section{Heating and cooling}
\label{s.heatingcooling}

The qualitative difference in the accretion rate history suggests that
only the strongly magnetized disk (model \texttt{Q}) can maintain the
sub-Eddington state. In this section we study the balance between
heating and cooling in the simulations.

Calculating the two is trivial for the simulations of accretion disks
in a shearing box where the gas is confined to the box and one may
study the evolution of the heat content in the same volume of gas for
a long time \citep{jiang+stability}. It is no longer straightforward in global
simulations such as discussed in this work. The reason is that the gas
does not stay in one location but moves in a turbulent way inward. 

It may seem that following the same parcel of gas along its way
towards the BH is the best approach to calculate heating and cooling in
a global simulation and we apply it below (Section~\ref{s.lagrangian}). However,
the heating and cooling rates depend on the location, and, for
simulations with limited range of inflow/outflow equilibrium, one
can follow the gas no longer than its inflow time.

Another approach is to study the gas properties at fixed radius. In
this way one may expect fixed heating and cooling rates, but one also
has to deal with different parcels of gas crossing a given radius at
a given time. Only if the system is in equilibrium, i.e., the average accretion
rate and disk properties do not change with time, one may say that at
fixed radius one probes different random (turbulent) realizations of
the same averaged state. We apply this method below as well
(Section~\ref{s.eulerian}).

To study the heating and cooling rates we define a ``box'' enclosed by
two radial boundaries and the vertical boundary at the polar angle 
$\theta=55^\circ$ (simulation \texttt{Q}) and $\theta=65^\circ$ (\texttt{D}), corresponding to the approximate locations of the
photosphere in the respective simulations (see the topmost panel
of Fig.~\ref{f.phisli} for the photosphere location in model \texttt{Q}), and covering
the full azimuthal extent of a given simulation. The cooling rate is
defined as the rate at which radiative flux crosses the vertical
boundaries of the box. The heating rate is estimated by calculating
the rate at which gas transfers energy to the radiation field, i.e.,
by integrating the time component of the radiative four-fource, $G_t$
(turbulent dissipation increases internal energy of the gas, but gas
immediately releases photons to maintain local thermal 
equilibrium\footnote{What would not be the case for optically thin
  disks.}).

\subsection{Eulerian picture}

\label{s.eulerian}
\begin{figure}
 \includegraphics[width=1.05\columnwidth]{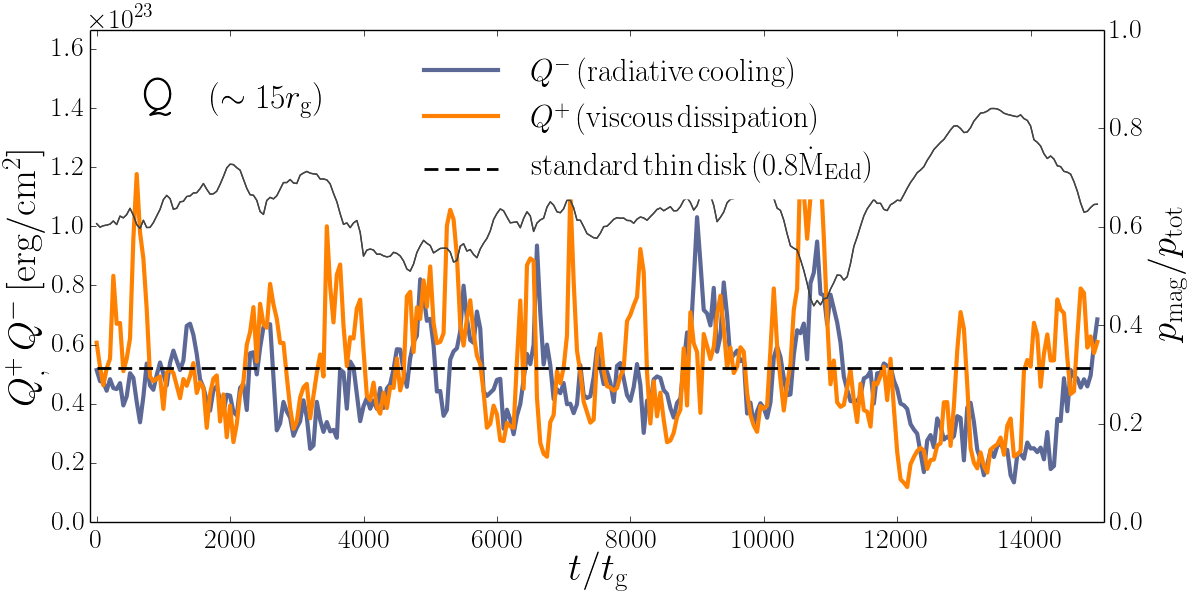}\hspace{-.5cm}
 \includegraphics[width=1.05\columnwidth]{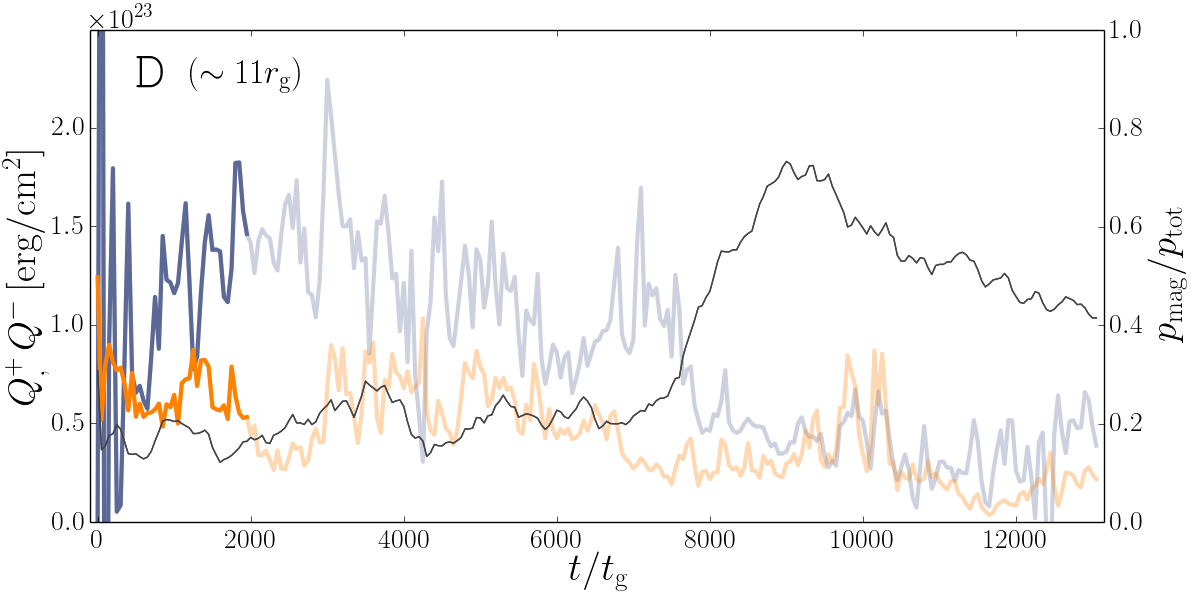}
 \caption{Heating ($Q^+$, orange lines) and cooling ($Q^-$, blue
   lines) as a function of time calculated at $r\approx 15\rg$ and $r\approx 11\rg$ for
   the stable simulation $\texttt{Q}$ (top) and the unstable
   simulation $\texttt{D}$ (bottom panel), respectively. The solid
   black line shows the magnetization parameter $\beta'=p_{\rm
     mag}/p_{\rm tot}$. The semi-transparent section of lines in the
   bottom panel reflect the period when the MRI was under-resolved at
   the equatorial plane.
  }
 \label{f.qplusqminus_vstime}
\end{figure}

\begin{figure*}
 \includegraphics[width=1.05\columnwidth]{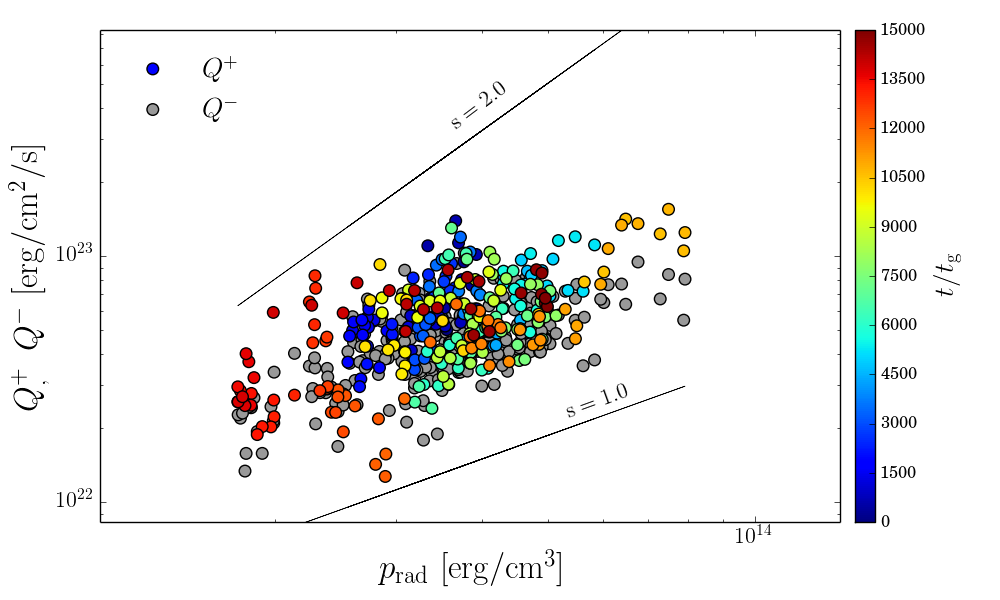}\hspace{-.5cm}
 \includegraphics[width=1.05\columnwidth]{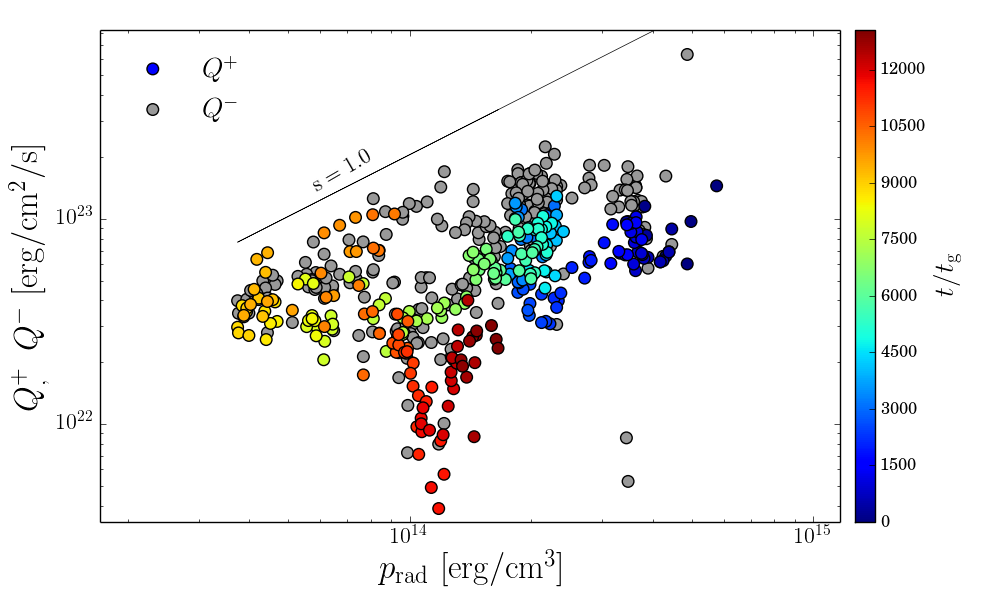}
 \caption{Heating ($Q^+$, color markers) and cooling ($Q^-$, grey markers) rates at $r\approx 15\rg$ for
   the stable simulation $\texttt{Q}$ (left panel) and the unstable
   simulation $\texttt{D}$ (right panel) as a function of the average radiation
   pressure, $p_{\rm rad}$. The values were extracted near radius
   $r=15\rg$ and $r=11\rg$ for models  $\texttt{Q}$ and  $\texttt{D}$,
   respectively. Colors of the markers
   denote the magnetic to total pressure ratio.
  }
 \label{f.qplusqminus}
\end{figure*}

We start studying heating and cooling by calculating the respective rates
at a fixed location. The radial boundaries of the boxes were set at
$r=14\rg$ and $16\rg$ for simulation \texttt{Q} and at $r=10\rg$ and
$11\rg$ for simulation \texttt{D} (the choice was motivated by the
condition that the boxes are inside the inflow/outflow equilibrium
region and far enough from the BH so that the effect of radial
advection of heat is negligible).

Figure~\ref{f.qplusqminus_vstime} shows the heating and cooling rates
calculated at these locations for model \texttt{Q} (top) and
\texttt{D} (bottom panel) as a function of time. For the former, the cooling rate follows
the viscous heating rate closely throughout the simulation and on
average corresponds to the rate predicted by the standard thin disk
model for accretion rate $0.8\Medd$. The magnetic contribution to
the total pressure stays in this case at the level of
$60\%$. The behavior of the weakly magnetized simulation \texttt{D}
(bottom panel) is
qualitatively diffferent. From the beginning the cooling rate
significantly exceeds the heating rate. As a result, the disk cools down, 
collapses towards the equatorial plane, and the MRI becomes
under-resolved at the equatorial plane for $t\gtrsim2000\tg$ (the
corresponding $Q^\theta$ parameter drops below $10$, see
\cite{hawley+13}). Because of cooling dominating over heating, the
thermal and radiation pressures keep decreasing, and ultimately (for
$t\gtrsim 8000\tg$), the magnetic pressure dominates the total
pressure. Whether such a
state can be maintained by itself will have to be verified numerically
in a separate, higher resolution, study.

Figure~\ref{f.qplusqminus} shows the heating/cooling rates as a
function of the (averaged over volume) radiation pressure for
the stable, strongly magnetized model \texttt{Q} (left panel) and the
collapsing model \texttt{D} (right panel). Two sets of markers are
plotted. The grey markers show the radiative cooling rates. The other
set of markers shows the heating rates and their colors denote the time
when they were calculated. For the magnetically supported disk (left
panel) the heating and cooling rates virtually overlap each other
proving the thermal equilibrium of the disk. There is no clear
temporal 
trend, i.e., the radiation pressure stays close to $3-4\times
10^{14} \,\rm erg/cm^3$ at all times. It is also evident that the rate of
radiative cooling is proportional to the radiation pressure,
consistent with the diffusive radiation transfer. Similar
dependence is shown by the rate of viscous dissipation,
in reasonable agreement with Eq.~\ref{e.dQdp}, which for $\beta'=0.6$ predicts
thermally stable 
slope of $s=0.8$.

The right panel shows the same quantities for simulation
\texttt{D}. In this case there is no equilibrium state -- the
accretion rate changes significantly with time and the gas crossing
given box is not expected to show consistent properties. This is
indeed the case. As noticed before, the cooling rate exceeds the
heating rate as long as the magnetization is low, i.e.,
$\beta'\lesssim 0.5$. The radiation pressure initially drops down
consistently with time showing once again that the disk was out of
thermal equilibrium. It stabilizes around $10^{14}\,\rm erg/cm^3$ only
in the late ($t\gtrsim 11000\tg$), strongly magnetized, but
underresolved stage.

\subsection{Lagrangian picture}
\label{s.lagrangian}

We now study the heating/cooling balance by following the
gas on its way towards the BH inside the inflow/outflow
equilibrium region. Because such an equilibrium has not been reached for
simulation \texttt{D}, we limit ourselves here to 
the stable simulation \texttt{Q}. We first approximate the average radial
velocity of the flow between radii $r=6\rg$ and $20\rg$ and get $v_{\texttt{Q}}=
-3(r/\rg)^{-2.5}$. Then, we ``drop'' a box, initially located between
$r=19\rg$ and $20\rg$, and calculate the heating/cooling rates on its
way towards the BH.

\begin{figure}
 \includegraphics[width=1.05\columnwidth]{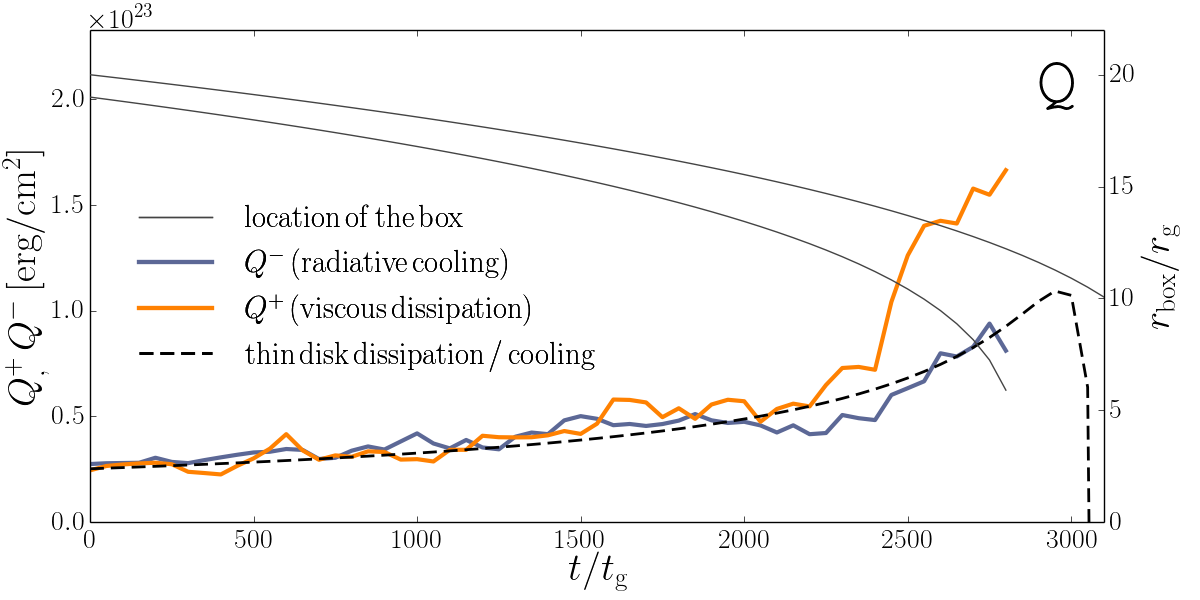}
 \caption{ Heating (orange) and cooling rates (blue lines) in a box
   infalling with the average radial velocity of the gas for
   simulation \texttt{Q}. The radial boundaries of the box were
   initially located at $r=19\rg$ and $20\rg$. The thin black lines
   show their location as a function of time.  }
 \label{f.Qsvstime}
\end{figure}

Figure~\ref{f.Qsvstime} shows the radiative cooling (blue
line) and viscous heating (orange line) rates as a function of time
and the corresponding location of the box (denoted with the black
line labeled on the right) for the stable simulation \texttt{Q}.
 The profiles were obtained by averaging the
properties of five separate boxes ``dropped'' at different times.
The dashed line shows the emission profile
predicted by the thin disk model for $0.8\Medd$ at the location
corresponding to the center of the box.

The simulated disk shows exact balance between heating
ang cooling in the initial stage of the box infall -- this indicates
that the disk is radiatively efficient, i.e., all the generated heat
is taken away from the disk by radiation. At the same time the disk
neither cools nor heats up, but maintains the equilibrium state it
has reached. When the inner boundary of the box is located inside
$r\approx 11\rg$ (what happens roughly after $t=2200\rg$), the rate of
viscous heating starts to exceed the rate of radiative cooling
(calculated by integrating radiative flux over the vertical
boundaries). However, most of the excess heat is ultimately deposited on the BH as a
result of both strong collimation of the radiation  (which
was emitted in gas moving with relativistic velocity towards the BH
and crosses the inner radial box boundary, instead of escaping vertically)
and photon trapping. The rate of radiative cooling, i.e., the rate at
which radiation escapes from the system, follows closely the
prediction of the standard model (denoted with the dashed line).

\section{Collapsing, weakly-magnetized disk}
\label{s.collapse}

\begin{figure}
 \includegraphics[width=1.05\columnwidth]{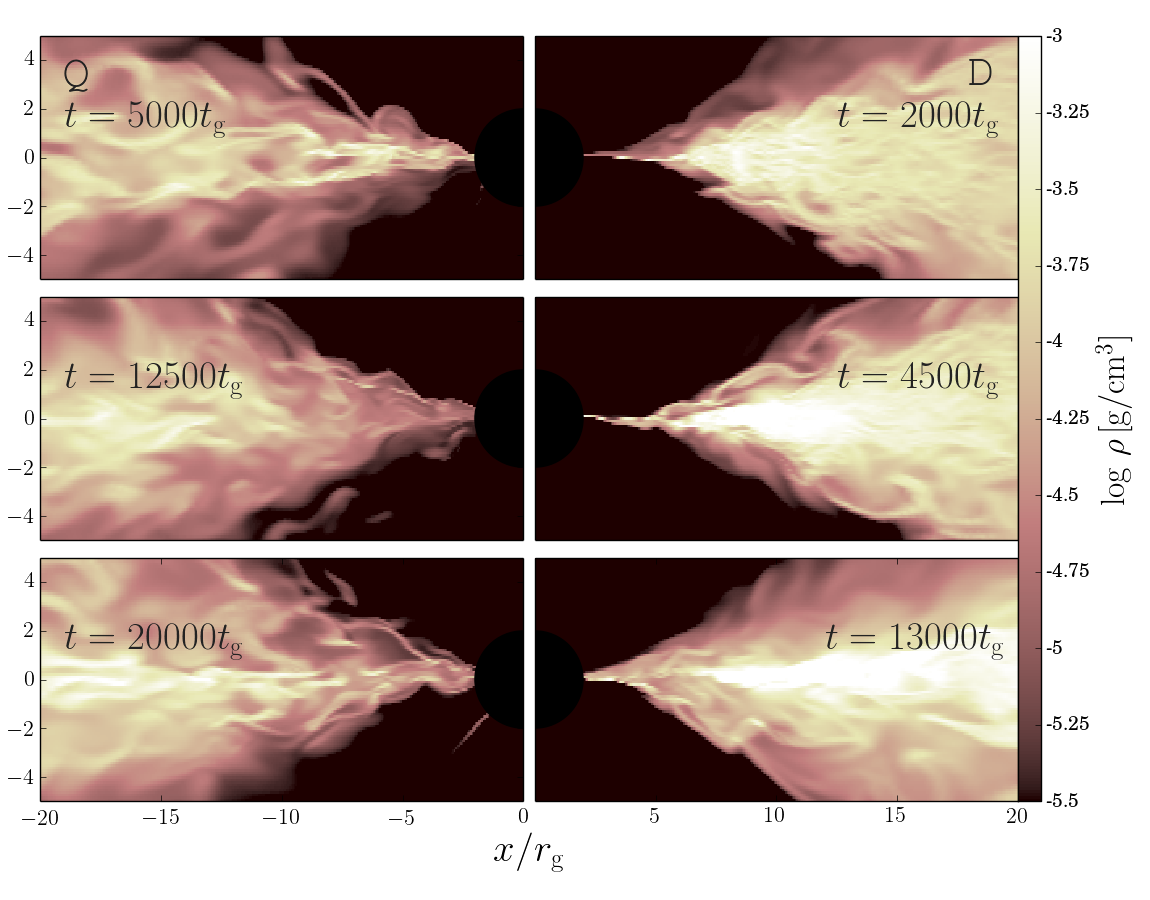}\vspace{-.2cm}
  \caption{Time evolution of the strongly (model
    \texttt{Q}, left panels) and weakly (model \texttt{D}, right
    panels) magnetized disks. Only the magnetically supported disk retains the
    equilibrium state. The weakly magnetized one cools down, collapses
  towards the equatorial plane and leads to under-resolving the MRI.}
 \label{f.collapse}
\end{figure}

Figure~\ref{f.collapse} compares the evolution of the two models over time. The
left set of panels shows the magnetically supported disk. It maintains
its thickness and properties throughout the simulation. The right
panels reflect the evolution of the weakly magnetized disk which
changes its properties on a relatively short (thermal) timescale. Because of the
excess of cooling over heating, the disk loses its radiation pressure
support and gas collapses towards the equatorial plane. The density
and the amount of gas in
that region increase, and the MRI becomes under-resolved.

Because of the cooling rate initially significantly exceeding the
heating rate, the radiative content of the collapsing disk decreases
and the magnetic field (dominated by the toroidal component 
preserved during the collapse) becomes more and more significant \citep{machida+06}. We
indeed observe that the latest stages of the evolution of model
\texttt{Q} become magnetic pressure supported
(Fig.~\ref{f.qplusqminus_vstime}). However, limited resolution of our
simulation does not allow us to study this state in detail. It is an open
question whethere the disk manages to stay in such strongly magnetized
(and therefore stable) state, despite the fact that there was hardly any net radial
flux of magnetic field before the collapse, or rather the limit-cycle
behavior \citep{lasotapelat-91,szuszkiewiczmiller-98} of the accretion flow leading to strongly varying accretion
rate (and luminosity) on timescales of the order of hours ($1h\approx
10^8\tg$) occurs.

\section{Properties of the stable solution}
\label{s.properties}

\begin{figure}
 \includegraphics[width=1.05\columnwidth]{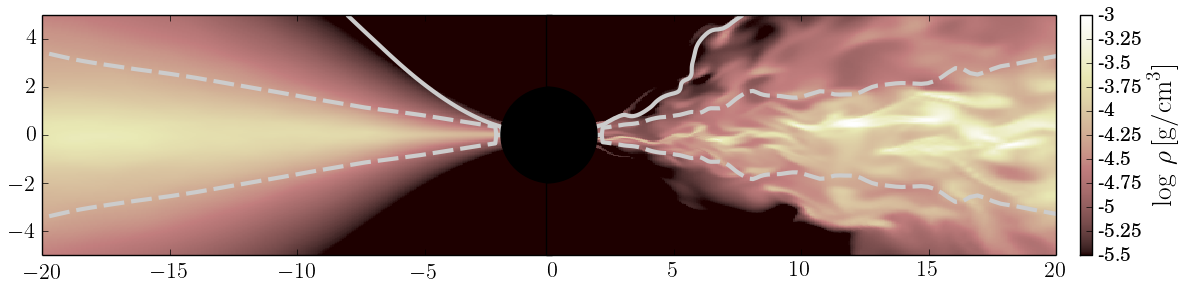}\vspace{-.2cm}
 \includegraphics[width=1.05\columnwidth]{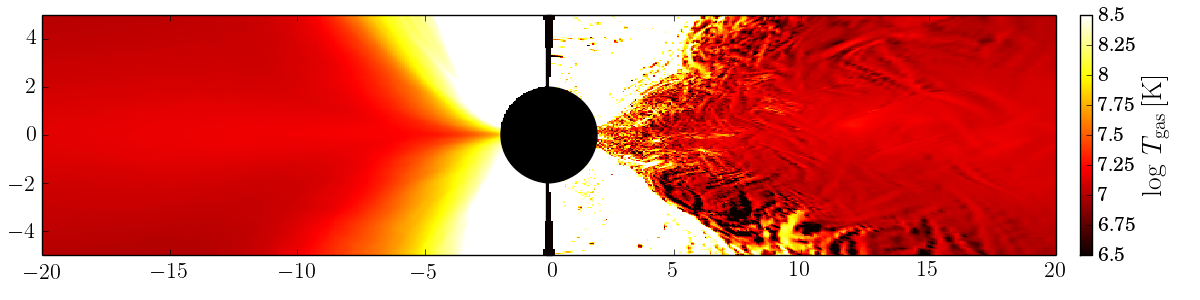}\vspace{-.2cm}
 \includegraphics[width=1.05\columnwidth]{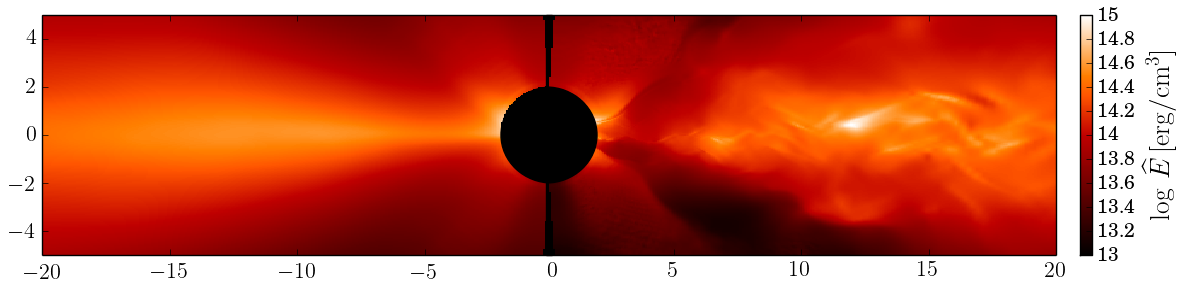}\vspace{-.2cm}
 \includegraphics[width=1.05\columnwidth]{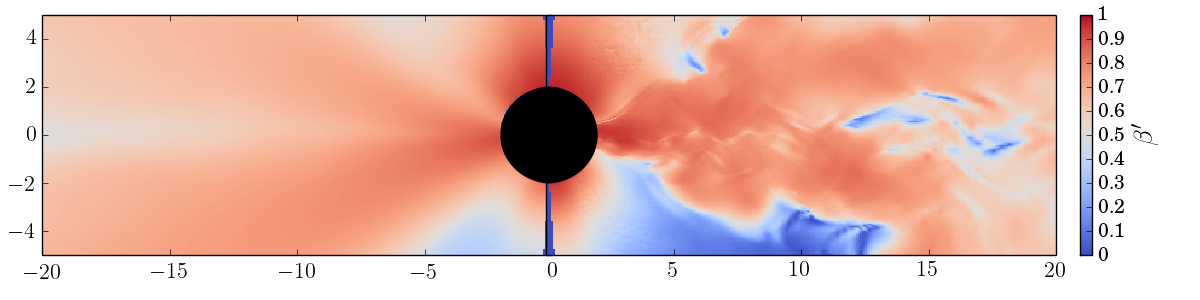}\vspace{-.2cm}
 \includegraphics[width=1.05\columnwidth]{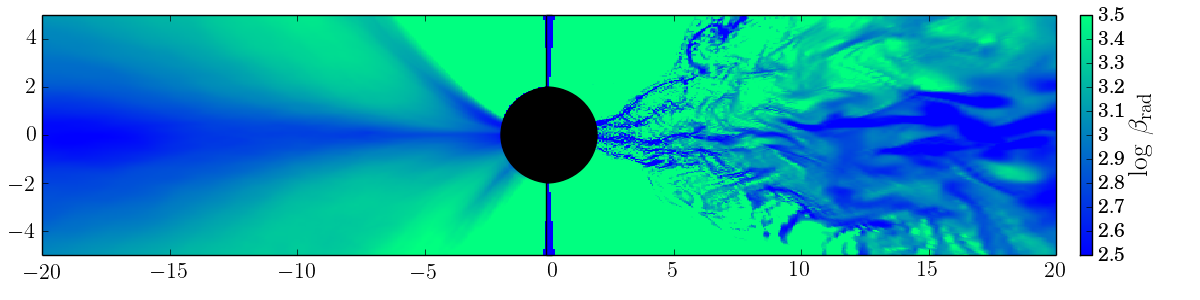}\vspace{-.2cm}
 \includegraphics[width=1.05\columnwidth]{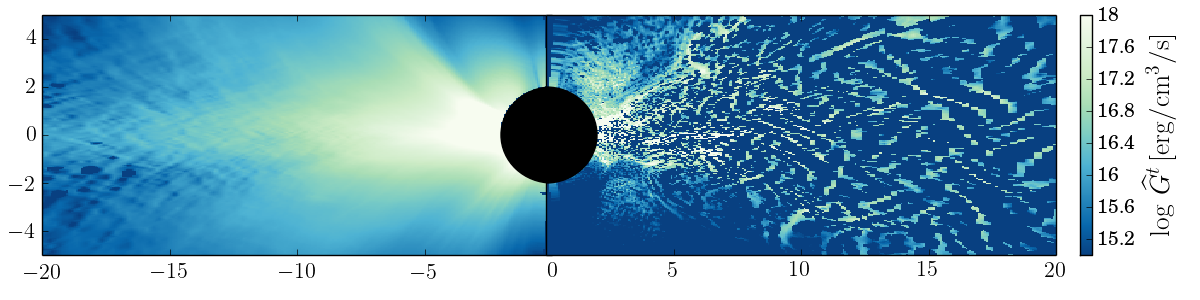}\vspace{-.2cm}
 \includegraphics[width=1.05\columnwidth]{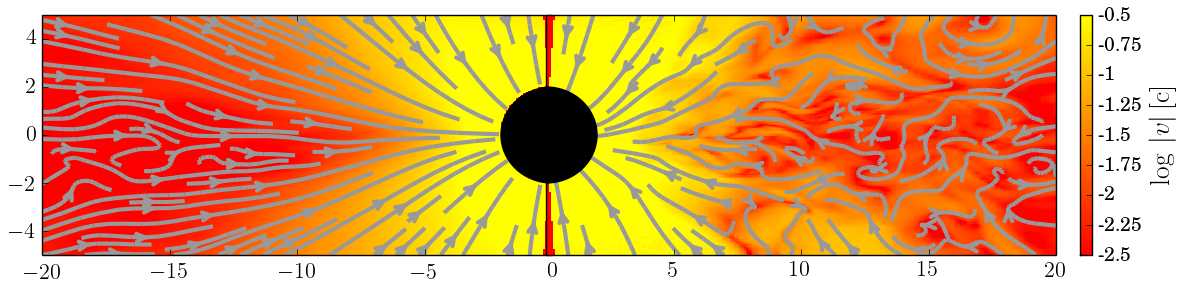}\vspace{-.2cm}
 \includegraphics[width=1.05\columnwidth]{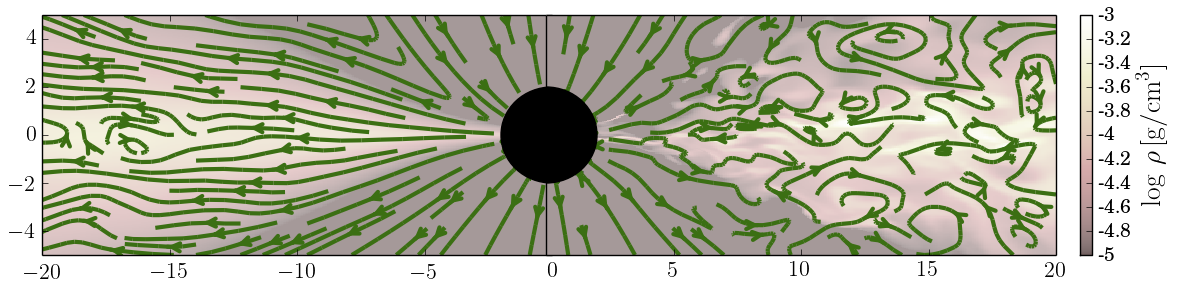}\vspace{-.2cm}
 \includegraphics[width=1.05\columnwidth]{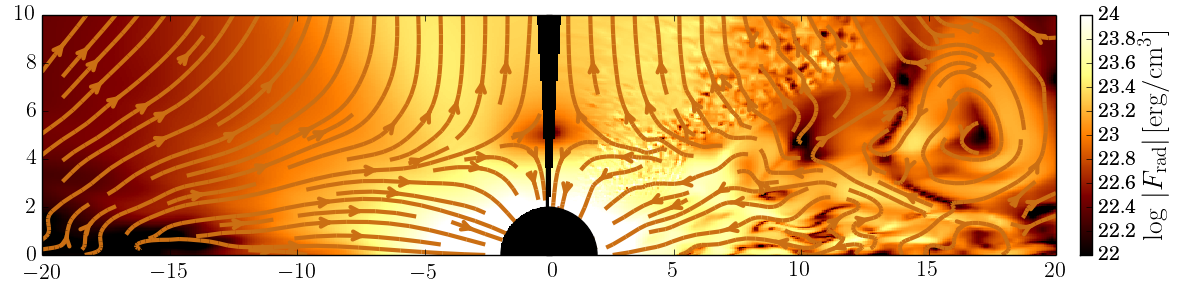}
  \caption{Comparison of time and azimuth averaged (left panels) and
    snapshot data (right panels) from the stable model \texttt{Q} reflecting, from top to bottom, gas
    density, $\rho$, gas temperature, $T_{\rm gas}$, comoving radiative
    energy density, $\widehat E$, magnetic to total pressure ratio,
    $\beta'$, radiation to gas pressure ratio, $\beta_{\rm
      rad}$,
    heating rate, $\widehat G^t$, magnitude and direction of gas
    velocity, $v$, poloidal magnetic field, $B^p$ (plotted on top of gas
    density), and magnitude and direction fo radiative flux, $F_{\rm
      rad}$. The solid and dashed lines in the topmost panel show the
    locations of the photosphere and density scale-height, respectively. }
 \label{f.phisli}
\end{figure}

Having identified the crucial role of magnetic field in stabilizing
disks at sub-Eddington accretion rates, we now look more closely at
the properties of the stable solution. We note, however, that the
solution we obtained is not unique. We suspect that the exact
properties of the disk will depend on the large-scale properties of
the magnetic field. Our simulation started with an arbitrary magnetic
field setup and evolved into a strongly magnetized state which proves
the concept, but does not have to be typical for the magnetically supported mode of accretion.
Further numerical work is necessary to get more understanding of such
accretion flows.

Figure~\ref{f.phisli} presents a number of disk properties on the
poloidal plane basing on the
averaged (left) and snapshot (right halves of the panels) data. The
first panel from the top shows the distribution of gas density, $\rho$ (colors),
together with the photosphere location (with respect to the total
opacity, solid lines) and
the density scaleheight, $h=\sqrt{\int \rho z^2 {\rm d}z/\int \rho
  {\rm d}z}$ (dashed lines). The gas is clearly turbulent and only
moderately concentrated at the equatorial plane. Although the
scaleheight is not large ($h/r\approx 0.15$), the photosphere is
quite far from the equatorial plane. Such a density distribution
results from the fact that the magnetic pressure
provides support against vertical gravity and prevents
significant compression of gas at the equatorial plane.

The second panel shows the gas temperature, $T_{\rm gas}$. Near the equatorial plane,
the gas is roughly at $10^{7.5}K$ in the whole inner region. Much
higher temperatures are near the axis where the density is the lowest,
and magnetization is the
largest. The distribution of temperature shows a slight
assymetry caused by the assymetric distribution of net magnetic field
(compare Fig.~\ref{f.Bfield}).

The third panel shows the energy density in the radiation field
measured in the comoving frame of gas, $\widehat E$ -- a quantity that in optically thick gas
reflects distribution of the radiation pressure 

The fourth panel shows the magnetic to total pressure ratio,
$\beta'$. The whole inner region is dominated by the
magnetic pressure. Even at the equatorial plane, where the radiation
pressure is largest, magnetic pressure
dominates the pressure balace. Near the scaleheight, but still below
the photosphere, the magnetic pressure contributes to 70-80\% of the
total pressure. Once again, the assymetry in the polar region
is visible.

The fifth panel shows the radiation to gas pressure ratio, $\beta_{\rm
  rad}$. The radiation pressure itself overwhelmes the thermal
pressure of  gas by roughly three orders of magnitude in the whole
domain.

The sixth panel shows the viscous heating rate, equal to the photon
generation rate, $\widehat G^t$. On average, it is positive and
reflects energy transfer from heated gas to the radiation field. The
dissipation takes place throughout the disk volume, and not only at
the equatorial plane, with significant amount of energy released just
below the photosphere and close to the BH horizon.

The seventh panel reflects the velocity field of the disk. On average,
the gas approaches the BH at all polar angles, even in the polar
region (what would presumably change if the BH was rotating and could
produce, even moderate, jet). The gas moves on average with the lowest velocity near the
equatorial plane. The surface layers fall on the BH with noticeably
larger velocity. As a result, the accretion is not limited to the most dense
equatorial region, but the surface layers bring inward comparable amount of
gas.

The eight panel (second from the bottom) shows the average and
temporal magnetic field. The average properties have alredy been
discussed in context of Fig.~\ref{f.Bfield}. The magnetic field is
not symmetric with respect to the equatorial plane -- magnetic
pressure is higher in the upper polar funnel and the net radial flux
in most of the disk is positive, causing significant growth of the
magnetic field and ultimately leading to the magnetically supported
state.

The bottommost panel in Fig.~\ref{f.phisli} shows the properties of
the radiation field. Colors denote the magnitude of the local
radiation flux and vectors show its direction. For $r\gtrsim 10\rg$
the radiation diffuses out of the disk in the direction roughly
perpendicular to the photosphere. Inside that radius, there is
significant component of radiation flux pointing towards the BH --
these are the photons which are either trapped in the optically thick
gas or were emitted, and therefore collimated, by gas
rushing towards the horizon. The radiation flux has the largest magnitude
inside the innermost stable orbit (ISCO) due to the strong
dissipation there (compare the sixth panel), but all that radiation
ends up ultimately under the horizon.

\section{Luminosity and energy flows in the stable solution}
\label{s.energetics}

We now briefly discuss energy flow in the magnetically supported,
sub-Eddington disk, basing on previously discussed in detail model
\texttt{Q}. We will follow closely the approach developed in
\cite{sadowski+enfluxes}. The total flux of energy is decomposed into
the binding, radiative and viscous fluxes. Binding energy is
transported with the gas which approaches the BH on roughly Keplerian orbits and
liberates its own binding energy (the sum of gravitational and
kinetic energies). Radiative energy is transported
either outward by free streaming photons escaping the system, or
inward by
photons trapped in optically thick gas or collimated towards the BH. Viscous energy transport results from{
the effective viscosity transferring rotational energy between
adjacent disk annuli. In a steady state, the three sum up to
the total, independent of radius, energy extracted from the system. In
the standard thin disk model, this total energy (or luminosity) 
equals to the (minus) binding energy of the gas crossing the horizon and, at the
same time, to the radiative luminosity of the system as seen from
infinity, i.e., to $5.7\%\,\dot Mc^2$ for a non-rotating BH ($5.7\,\%$ is
the radiative efficiency of the system).

Figure~\ref{f.enfluxes} shows such a decomposition of the total energy
flux for simulation \texttt{Q} based on time and azimuth averaged data. The total energy, denoted by the thick
solid red line, is constant to $10\%$ within $r\approx 20\rg$ (the
departure from a constant value suggests that the properties of the disk
change a bit with time). The total luminosity of the system is very
close to the thin disk luminosity (denoted by the red dashed line),
and equals $5.5\pm0.5\%\,\dot M c^2$.

The binding energy (blue solid line) is also close to the thin disk
counterpart (blue dashed line) outside the innermost stable orbit
($r_{\rm ISCO}=6\rg$). This fact reflects almost Keplerian angular
momentum of the gas in the discussed simulation and relatively
insignificant radial velocity contribution. The behavior is, however,
qualitatively different inside the ISCO. While in the standard thin disk model
there is no torque acting on the gas inside that
radius \citep{paczynski-00}\footnote{\cite{paczynski-00} has shown that
  the torque at the inner edge dissapears if only angular momentum is
  conserved, disk is thin ($h/r\ll 1$), and the viscosity parameter is
  small ($\alpha\ll 1$). In our solution the last condition is not
  satisfied, with $\alpha \approx 0.3$ at ISCO and reaching $0.9$ near
  $r=3\rg$.} and binding
energy is constant in the plunging region, the numerical solution
shows that gas becomes more and more bound after crossing ISCO and,
as a result, more and more energy is transported as binding energy. This means
that significant stresses act inside ISCO, what should not be surprising having
in mind how strongly magnetized the gas is (see \cite{noble+10}, but
also \cite{shafee+08,penna+10,kulkarni+11}).

The profile of radiative luminosity is also very close to the thin
disk profile (compare solid and dashed orange lines). At radius
$r=15\rg$ radiative flux carries out $\sim 1\%\,\dot Mc^2$, only
ca. $10\%$ less than the standard model predicts. The behavior in the
innermost region is, however, again very different. According to the standard thin
disk model, no radiation crosses the ISCO or is generated inside that
radius. In this numerical simulation, on the contrary, significant
dissipation takes place in the innermost region. Nevertheless, most of the
radiation emitted there ends up inside
the BH because it is either trapped in the flow or is strongly
collimated towards the BH (compare the bottommost panel of Fig.~\ref{f.phisli}). Photons
crossing the horizon carry roughly $2\%\,\dot M c^2$.

Finally, the viscous flux of energy once again agrees well in the outer region with the
predictions of the analytical thin disk modeling
but disagrees in the innermost
($r\lesssim 10\rg$) region where noticeable viscous energy transport
takes place, in disagreement with the standard modeling.

To sum up, the simulated magnetically supported thin disk shows
significant dissipation and stress inside the ISCO, in the
so-called plunging region, inconsistent with the
standard model. However, somewhat surprisingly, the energetics of the
disk far from the BH agrees very well with the standard modeling. The
extra energy extracted in the innermost region does not have a chance
of being observed from infinity\footnote{Similar conclusion was
  reached by \cite{zhu+eye} who studied radiation from artificially
  cooled thin MHD disks.}. Whether this is true also for thinner, less
optically thick disks, corresponding to lower accretion rates, has to
be verified numerically in the future.

\begin{figure}
 \includegraphics[width=1.\columnwidth]{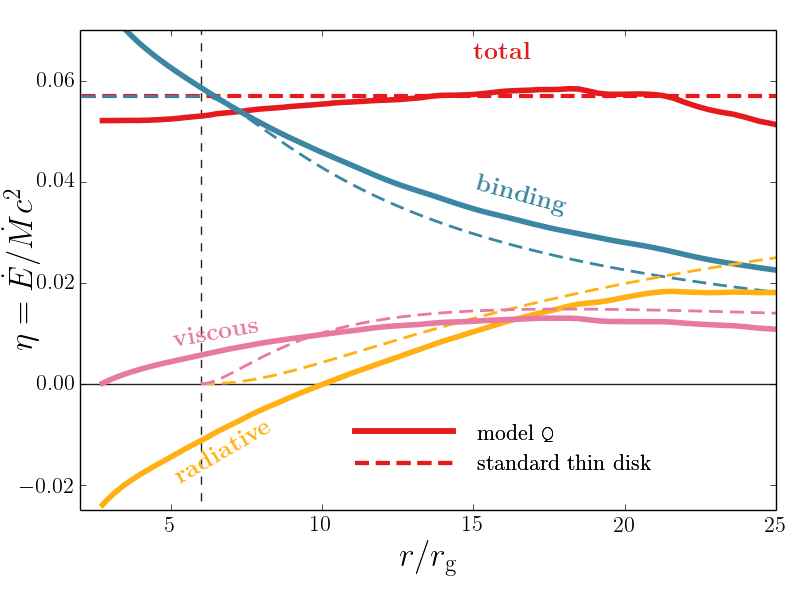}
 \caption{
Energy flow through the stable simulation \texttt{Q} (solid lines) compared to
the thin-disk solution (dashed lines). 
  }
 \label{f.enfluxes}
\end{figure}

\section{Discussion}
\label{s.discussion}

\subsection{Net flux of magnetic field}

We have shown that thin and radiatively efficient accretion disks 
are thermally stable when dominated by magnetic pressure. The necessary condition for the stability is
strong magnetic support which is provided mainly by the azimuthal
component of the field. In the case of our stable simulation (model
\texttt{Q}) such strong azimuthal field developed as a result of
non-zero radial flux of magnetic field inside the disk 
Are astrophysical accretion disks likely to have significant net
radial flux? 

It is tempting to expect that the origin of such a large-scale field
is the advection of the field by the gas in the accretion disk from
the interstellar medium or a companion star. If only such a field is
not perfectly dipolar (or uneven-polar) one may expect that the
asymmetry leads to effectively a non-zero radial flux in the
bulk of the disk. However, whether or not the large-scale field can be
advected into the innermost region depends on the balance between the
advection and diffusion of the field. Standard geometrically thin
disks are unlikely to drag significant amount of magnetic field on to
the BH \citep{lubow+94,ghosh+97}. However, as
\cite{guletogilvie-12,guletogilvie-13} found, if one takes into
account the vertical structure of the disk, the advection of magnetic
flux can be much faster than found originally. Indeed, it may be the case for
the strongly magnetized solutions discussed here which not only show much larger
radial velocities than the standard model predicts, but also the
velocities increase significantly towards the surface of the disk.

There is also another way of obtaining an accretion disk
dominated by magnetic pressure which, on the contrary, does not
require (at least initially)
efficient advection of magnetic field. As \cite{machida+06} pointed out, when an
X-ray binary undergoes the transition from the low/hard state
(corresponding to low accretion rates and low optical depths) to
the high/soft state (at larger accretion rates), the disk starts to
cool and shrinks in the vertical direction while almost conserving the
toroidal magnetic flux (we see such behavior in the collapse of model \texttt{D}). Such a process leads to a cooler, magnetically
supported, quasi-steady state. \cite{machida+06} have shown that such
a configuration can exist for at least the thermal time. Whether it can
survive for much longer times, without relaxing to the lower
magnetization state implied by local, zero-flux numerical studies, has
still to be verified \citep[note recent work by][]{wielgus+15}.

\subsection{Observables}

The magnetically supported thin disks, an example of which has been
presented in this work, differ from the standard model of thin disks
\citep{ss73} in two aspects. Firstly, they are supported by
magnetic pressure, not the radiation pressure, although the two have
similar magnitudes. This fact results in a disk thicker than standard
one at the same radius and accretion rate. The effective color
temperature is also higher because the photosphere is now located in
the strongly magnetized, hot region where photons are  efficiently
up-scattered by hot electrons. Secondly, the high magnetization
implies significant stresses taking place at and inside ISCO. As a
result, radiation is efficiently generated in the plunging region,
in contrary to the assumptions of the standard model. However,
for the slightly sub-Eddington disk presented in this work, all this
excess of generated radiation is advected on the BH, and the
luminosity of the system, as seen from the distance, very well agrees
with the efficiency of the standard disks. 

What would such a magnetized thin disk system look like? Because the
extra radiation generated by dissipation in the plunging region does
not escape, one may expect similar luminosities to the predicted by
the standard model. However, the spectrum may look different. Thicker
disk would imply larger color temperature and harder spectrum (what
would bring continuum-based BH spin estimates down). Also a
stronger inclination dependence may be expected. Better answers will
be provided once detailed frequency- and angle-resolved calculations
of the radiation transfer based on numerical simulations like ours
are performed.

\section{Summary}
\label{s.summary}

We have performed two three-dimensional, global, general relativistic,
radiative simulations of sub-Eddington accretion disks. The
simulations were initialized with different configurations of the
magnetic field. One of them evolved into a magnetically supported
state in thermal equilibrium with strong magnetic field maintained by a non-zero radial
magnetic flux in the bulk of the disk. The other did not develop
strong magnetic field, showed consisted excess of cooling over heating
and continued collapsing towards the equatorial plane. We summarize
our conclusions as following:

\begin{enumerate}
\item \textit{Thermal stability:} -- We have shown that magnetic
  pressure dominated thin, sub-Eddington accretion disks can maintain
  thermal equilibrium, in contrast to thermally unstable disks
  dominated by radiation pressure.

\item \textit{Magnetization:} -- The strong magnetization required for
  stabilization of thin disks results either from the presence of
  a non-zero radial magnetic flux (what requires efficient advection
  of large-scale magnetic field by the disk), or from the collapse of
  previously thick disk during the hard to soft transition.

\item \textit{Standard model:} -- The magnetically supported thin
  disks are thicker than standard ones at the same radius and
  accretion rate. They also show significant dissipation inside
  ISCO. Radiation released in this region, however, does not escape
  but is advected on the BH. Whether this statement is true for lower
  accretion rates than simulated here ($0.8\Medd$) has still to be
  verified.

\item \textit{Luminosity:} -- Despite strong dissipation and torque
  inside ISCO, the efficiency of the simulated disk ($5.5\pm0.5\%$) turns
  out to be very close to the prediction of the standard model
  ($5.7\%$). The spectrum of the observed radiation, however, is
  likely to be harder due to the photosphere extending further away
  from the equatorial plane. Detailed spectral calculations of this
  mode of accretion will follow.

\end{enumerate}

\section{Acknowledgements}

The author thanks David Abarca for his contribution and Ramesh
Narayan, Jean-Pierre Lasota and Marek Abramowicz for helpful comments.
The author acknowledges support
for this work 
by NASA through Einstein Postdoctotral Fellowship number PF4-150126
awarded by the Chandra X-ray Center, which is operated by the
Smithsonian
Astrophysical Observatory for NASA under contract NAS8-0306These authors 0. 
The author acknowledges computational support from NSF via XSEDE resources
(grants TG-AST080026N and TG-AST150019), and
from NASA via the High-End Computing (HEC) Program
through the NASA Advanced Supercomputing (NAS) Division at Ames
Research Center. The author was supported in part by NSF grant
AST1312651 
and NASA grant TCAN NNX14AB47G.
The author also acknowledges support from the International Space Science Institute.
 
\bibliographystyle{mn2e}

{\small
\begin{thebibliography}{}



\bibitem[Abramowicz et al.(1988)]{abra88} Abramowicz, M. A., Czerny,
  B., Lasota, J.~P., \& Szuszkiewicz, E.\ 1988, \apj, 332, 646


\bibitem[Balbus 
\& Hawley(1991)]{balbushawley-mri} Balbus, S.~A., \& Hawley, J.~F.\ 1991, \apj, 376, 214 



\bibitem[Bai 
\& Stone(2013)]{xuening+13} Bai, X.-N., \& Stone, J.~M.\ 2013, \apj, 767, 30 

\bibitem[Begelman 
\& Pringle(2007)]{begelmanpringle-07} Begelman, M.~C., \& Pringle, J.~E.\ 2007, \mnras, 375, 1070 


\bibitem[Begelman 
\& Armitage(2014)]{begelmanarmitage-14} Begelman, M.~C., \& Armitage, P.~J.\ 2014, \apjl, 782, L18 



\bibitem[Ciesielski et 
al.(2012)]{ciesielski+12} Ciesielski, A., Wielgus, M., Klu{\'z}niak, W., et al.\ 2012, \aap, 538, A148 

\bibitem[Coriat et al.(2012)]{coriat+12} Coriat, M., Fender, 
R.~P., \& Dubus, G.\ 2012, \mnras, 424, 1991 

\bibitem[Davis et al.(2010)]{davis+10} Davis, S.~W., Stone, 
J.~M., \& Pessah, M.~E.\ 2010, \apj, 713, 52 


\bibitem[Ghosh 
\& Abramowicz(1997)]{ghosh+97} Ghosh, P., \& Abramowicz, M.~A.\ 1997, \mnras, 292, 887 

\bibitem[Guilet 
\& Ogilvie(2012)]{guletogilvie-12} Guilet, J., \& Ogilvie, G.~I.\ 2012, \mnras, 424, 2097 



\bibitem[Guilet 
\& Ogilvie(2013)]{guletogilvie-13} Guilet, J., \& Ogilvie, G.~I.\ 2013, \mnras, 430, 822 

\bibitem[Hawley et al.(2013)]{hawley+13} Hawley, J.~F., Richers, S.~A., Guan, X., \& Krolik, J.~H.\ 2013, \apj, 772, 102 


\bibitem[Hirose et al.(2009)]{hirose+09} Hirose, S., Krolik, 
J.~H., \& Blaes, O.\ 2009, \apj, 691, 16 


\bibitem[Janiuk 
\& Misra(2012)]{janiukmisra-12} Janiuk, A., \& Misra, R.\ 2012, \aap, 540, A114 

\bibitem[Jiang, Stone, 
\& Davis(2013)]{jiang+stability} Jiang, Y.-F., Stone, J.~M., \& Davis, S.~W.\ 2013, \apj, 778, 65 

\bibitem[Jiang et al.(2014)]{jiang+3dsim} Jiang, Y.-F., Stone, 
J.~M., \& Davis, S.~W.\ 2014, \apj, 796, 106 

\bibitem[Kulkarni et al.(2011)]{kulkarni+11} Kulkarni, A.~K., 
Penna, R.~F., Shcherbakov, R.~V., et al.\ 2011, \mnras, 414, 1183 


\bibitem[Lasota 
\& Pelat(1991)]{lasotapelat-91} Lasota, J.~P., \& Pelat, D.\ 1991, \aap, 249, 574 

\bibitem[Lasota(2001)]{lasota+instability} Lasota, J.-P.\ 2001, New
  Astronomy Review, 45, 
449 

\bibitem[Levermore(1984)]{levermore84} Levermore, C.~D.\ 1984, 
\jqsrt, 31, 149 

\bibitem[Li 
\& Begelman(2014)]{libegelman-14} Li, S.-L., \& Begelman, M.~C.\ 2014, \apj, 786, 6 

\bibitem[Lightman 
\& Eardley(1974)]{lightmaneardley-74} Lightman, A.~P., \& Eardley, D.~M.\ 1974, \apjl, 187, L1 


\bibitem[Lubow et al.(1994)]{lubow+94} Lubow, S.~H., Papaloizou, 
J.~C.~B., \& Pringle, J.~E.\ 1994, \mnras, 267, 235 


\bibitem[Maccarone(2003)]{maccarone+03} Maccarone, T.~J.\ 2003, \aap, 409, 697 

\bibitem[Machida et al.(2006)]{machida+06} Machida, M., Nakamura, 
K.~E., \& Matsumoto, R.\ 2006, \pasj, 58, 193 



\bibitem[McClintock 
\& Remillard(2006)]{mcclintock+06} McClintock, J.~E., \& Remillard, R.~A.\ 2006, Compact stellar X-ray sources, 39, 157 


\bibitem[McKinney et al.(2014)]{mckinney+harmrad} McKinney, J.~C., 
Tchekhovskoy, A., Sadowski, A., \& Narayan, R.\ 2014, \mnras, 441, 3177 


\bibitem[Noble et al.(2010)]{noble+10} Noble, S.~C., Krolik, 
J.~H., \& Hawley, J.~F.\ 2010, \apj, 711, 959 


\bibitem[Oda et al.(2009)]{oda+09} Oda, H., Machida, M., 
Nakamura, K.~E., \& Matsumoto, R.\ 2009, \apj, 697, 16 



\bibitem[Ohsuga 
\& Mineshige(2011)]{ohsuga11} Ohsuga, K., \& Mineshige, S.\ 2011, \apj, 736, 2 


\bibitem[Paczy{\'n}ski(2000)]{paczynski-00} Paczy{\'n}ski, B.\ 2000, 
arXiv:astro-ph/0004129 


\bibitem[Penna et al.(2010)]{penna+10} Penna, R.~F., McKinney, 
J.~C., Narayan, R., et al.\ 2010, \mnras, 408, 752 

\bibitem[Penna et al.(2013)]{penna+alpha} Penna, R.~F., S{\c 
a}dowski, A., Kulkarni, A.~K., \& Narayan, R.\ 2013, \mnras, 428, 2255 


\bibitem[Pessah 
\& Psaltis(2005)]{pessahpsaltis-05} Pessah, M.~E., \& Psaltis, D.\ 2005, \apj, 628, 879 

\bibitem[Piran(1978)]{piran-78} Piran, T.\ 1978, \apj, 221, 652 

\bibitem[Pringle(1976)]{pringle-76} Pringle, J.~E.\ 1976, \mnras, 
177, 65 


\bibitem[S{\c a}dowski et al.(2013a)]{sadowski+koral} S{\c a}dowski, 
A., Narayan, R., Tchekhovskoy, A., \& Zhu, Y.\ 2013a, \mnras, 429, 3533 




\bibitem[S{\c a}dowski et al.(2014a)]{sadowski+koral2} S{\c a}dowski, A., Narayan, R., McKinney, J.~C., \& Tchekhovskoy, A.\ 2014b, \mnras, 439, 503 

\bibitem[S{\c a}dowski et al.(2014b)]{sadowski+dynamo} S\k{a}dowski, A., Narayan, R., Tchekhovskoy, A., Abarca, D., Zhu, Y., \& McKinney J.~C. 2014b, in press

\bibitem[S{\c a}dowski et al.(2016)]{sadowski+enfluxes} S{\c a}dowski, 
A., Lasota, J.-P., Abramowicz, M.~A., 
\& Narayan, R.\ 2016, \mnras, 456, 3915 

\bibitem[Shakura \& Sunyaev(1973)]{ss73}Shakura, N. I., \& Sunyaev, R. A. 1973, A\&A, 24, 337

\bibitem[Shakura 
\& Sunyaev(1976)]{ss76} Shakura, N.~I., \& Sunyaev, R.~A.\ 1976, \mnras, 175, 613 

\bibitem[Shafee et al.(2008)]{shafee+08} Shafee, R., McKinney, 
J.~C., Narayan, R., et al.\ 2008, \apjl, 687, L25 

\bibitem[Shi et al.(2010)]{shi+10} Shi, J., Krolik, J.~H., 
\& Hirose, S.\ 2010, \apj, 708, 1716 


\bibitem[Szuszkiewicz 
\& Miller(1998)]{szuszkiewiczmiller-98} Szuszkiewicz, E., \& Miller, J.~C.\ 1998, \mnras, 298, 888 


\bibitem[Wielgus et al.(2015)]{wielgus+15} Wielgus, M., Fragile, 
P.~C., Wang, Z., \& Wilson, J.\ 2015, \mnras, 447, 3593 


\bibitem[Zhu et al.(2012)]{zhu+eye} Zhu, Y., Davis, S.~W., 
Narayan, R., et al.\ 2012, \mnras, 424, 2504 



\end{thebibliography}
}

\end{document}